\documentclass{elsart}

\usepackage{epsfig}

\usepackage{amssymb}

\begin{document}

\begin{frontmatter}

\title{Radiative instability of a relativistic electron beam moving
in a photonic crystal}

\author{V.G. Baryshevsky, A.A. Gurinovich}

\address{Research Institute for Nuclear Problems, Minsk, Belarus}

\begin{abstract}
The radiative instability of a beam moving in a photonic crystal
of finite dimensions is studied. The dispersion equation is
obtained. The law $\Gamma\sim\rho ^{1/\left( {s + 3} \right)}$ is
shown to be valid and caused by the mixing of the electromagnetic
field modes in the finite volume due to the periodic disturbance
from the photonic crystal.
\end{abstract}

\end{frontmatter}

Numerous works analyze the generation of induced radiation by
electron beams moving in spatially periodic media
\cite{1,2,3,4,5}. Thoroughly studied are the processes of wave
generation in a one-dimensional case, when the electron beam moves
along the axis of a corrugated waveguide (traveling-wave tube
(TWT), backward-wave tube (BWT)) or along the axis of the
undulator (volume free electron lasers, ubitrons). It was found
out that under the action of radiation, the electron beam with a
uniform density distribution becomes spatially modulated, i.e.,
the radiative instability of a beam emerges. The increment of the
electron beam instability is a most important quantity
characterizing the generation capability of the beam. The analysis
of all mechanisms of induced radiation generation by relativistic
beams in the Compton regime showed that the increment of beam
instability $\Gamma $ for a cold beam (i.e., a beam where all the
electrons have the same longitudinal velocity $\vec{u}$) follows
the law $\Gamma \sim \rho ^{1/3}$, where $\rho $ - is the beam
density \cite{2,3}.

In 1984 it was shown \cite{6} that induced X-ray radiation
produced by the electron beam passing through the crystal under
the conditions providing the coincidence of roots of the
dispersion equation, which describes the relation between the wave
vector $\vec {k}$ and the photon frequency $\omega $ in the
crystal, leads to the appearance of a new law for the increment of
the beam instability:  $\Gamma  \sim \rho ^{1/\left( {s + 3}
\right)}$, where $s$ is the number of additional waves appearing
due to diffraction of emitted X-ray quanta in the crystal. This
law of instability leads to a significant reduction of the laser
generation threshold for X-ray radiation in the crystal (according
to this law, the generation threshold for a LiH crystal can be
achieved when the beam current density is $10^{8}$~ A/cm$^{2}$ in
contrast to $10^{13}$~ A/cm$^{2}$ as required according to the
conventional law $\rho ^{1/3}$.

It was later shown in \cite{7,8} that the above described law is
valuid for all wavelength ranges of induced radiation produced by
electrons in spatially periodic media (diffraction gratings) and
for various types of nonlinear interaction of waves in both
natural and artificial (or frequently called  "photonic")
crystals. The results obtained in these works enable design and
development of a new type of Free Electron Lasers called the
Volume free Electron Lasers (VFEL) \cite{9,10}.

Theoretical description of the generation processes in a photonic
crystal placed inside the resonator was given in \cite{9,12,13}.
The first and most important step in describing the generation
process in VFELs (FELs and so on) is the analysis of the problem
of the electron beam instability in the resonator. The theoretical
study of the instability of electron beams moving in natural and
artificial (photonic) crystals was carried out for the ideal case
of an infinite medium (see the review in \cite{9} and
\cite{6,10,11,12,13}). The question arising in this regard is how
the finite dimensions of the photonic crystal placed inside the
resonator affect the law of electron beam instability. It is
known, for example, that the discrete structure of the modes in
waveguides and resonators is crucial for effective generation in
the microwave range \cite{2,3,4,5}.

In present paper the radiative instability of a beam moving in a
photonic crystal is studied. The dispersion equation describing
instability in this case is obtained. It is shown that the law
$\Gamma\sim\rho ^{1/\left( {s + 3} \right)}$ is also valid and
caused by the mixing of the electromagnetic field modes in the
finite volume due to the periodic disturbance from the photonic
crystal.

The system of equations describing generation of induced radiation
in photonic (and natural) crystals  can be obtained from Maxwell
equations:

\begin{eqnarray}
 rot\vec {H} = \frac{{1}}{{c}}\frac{{\partial \vec
{D}}}{{\partial t}} + \frac{{4\pi} }{{c}}\vec {j},\;rot\vec {E} =
- \frac{{1}}{{c}}\frac{{\partial \vec {H}}}{{\partial t}}, \\
 div\vec {D} = 4\pi \rho ,\;\frac{{\partial \rho}
}{{\partial t}} + div\vec {j} = 0,\nonumber
\label{eq1}
\end{eqnarray}

\noindent where $\vec {E}$ and  $\vec {H}$  are the strength of
the electric and the magnetic field, respectively;
 $\vec {j}$ è $\rho $ are the current and charge densities;
 $D_{i} \left( {\vec
{r},t^{\prime} } \right) = \smallint \varepsilon _{il} \left(
{\vec {r},t - t^{\prime} } \right)E_{l} \left( {\vec
{r},t^{\prime} } \right)dt^{\prime} $ or $D_{i} \left( {\vec
{r},\omega}  \right) = \varepsilon _{il} \left( {\vec {r},\omega}
\right)E_{l} \left( {\vec {r},\omega}  \right)$, where indices
$i,l = 1,2,3$ correspond to $x,y,z$;  $\varepsilon _{il} \left(
{\vec {r},\omega}  \right)$ is the dielectric permittivity tensor
of the photonic crystal.

The current and charge densities are defined as:

\begin{equation}
\label{eq2} \vec {j}\left( {\vec {r},t} \right) = e \sum_\alpha
{\vec {v}}_\alpha \delta(\vec {r} -\vec {r}_\alpha(t)), ~\rho(\vec
{r},t)=e \sum_\alpha \delta(\vec {r} -\vec {r}_\alpha(t)),
\end{equation}

\noindent where $e$ is the electron charge, $\mathop {\vec
{v}}_{\alpha}$ is the velocity of the electron with number $\alpha
$ in the electron beam,

\begin{equation}
\label{eq3} \frac{{d  {\vec {v}}_{\alpha} } }{{dt}} =
\frac{{e}}{{m\gamma _{\alpha} } }\left\{ {\vec {E}\left( { {\vec
{r}}_{\alpha} \left( {t} \right),t} \right) +
\frac{{1}}{{c}}\left[ {  {\vec {v}}_{\alpha} \left( {t} \right)
\times \vec {H}\left( {  {\vec {r}}_{\alpha} \left( {t} \right),t}
\right)} \right] - \frac{{  {\vec {v}}_{\alpha} } }{{c^{2}}}\left(
{  {\vec {v}}_{\alpha}  \left( {t} \right)\vec {E}\left( { {\vec
{r}}_{\alpha}  \left( {t} \right),t} \right)} \right)} \right\},
\end{equation}

\noindent where $\gamma _{\alpha}  = \left( {1 -
{\textstyle{{v_{\alpha} ^{2}}  \over {c^{2}}}}} \right)^{ -
{\textstyle{{1} \over {2}}}}$ is the Lorentz factor, $\vec
{E}\left( { {\vec {r}}_{\alpha}  \left( {t} \right),t} \right)$
($\vec {H}\left( {  {\vec {r}}_{\alpha}  \left( {t} \right),t}
\right)$) is the electric (magnetic) field strength at point $
{\vec {r}}_{\alpha}  $, where the electron with number $\alpha $
is located. Note that equation (\ref{eq3}) can be written as
follows \cite{14}:

\begin{equation}
\label{eq4} \frac{{d  {\vec {p}}_{\alpha} } }{{dt}} =
m\frac{{d\gamma _{\alpha}  v_{\alpha} } }{{dt}} = e\left\{ {\vec
{E}\left( {  {\vec {r}}_{\alpha}  \left( {t} \right),t} \right) +
\frac{{1}}{{c}}\left[ {  {\vec {v}}_{\alpha} \left( {t} \right)
\times \vec {H}\left( {  {\vec {r}}_{\alpha} \left( {t} \right),t}
\right)} \right]} \right\},
\end{equation}

\noindent
where $p_{\alpha}  $ is the particle momentum.

From equations (\ref{eq1}) one can obtain

\begin{equation}
\label{eq5}
 - \Delta \vec {E} + \vec {\nabla} \left( {\vec {\nabla} \vec {E}} \right) +
\frac{{1}}{{c^{2}}}\frac{{\partial ^{2}\vec {D}}}{{\partial t^{2}}} = -
\frac{{4\pi} }{{c^{2}}}\frac{{\partial \vec {j}}}{{\partial t}}.
\end{equation}

The dielectric permittivity tensor can be presented in the form $\hat
{\varepsilon} \left( {\vec {r}} \right) = 1 + \hat {\chi} \left( {\vec {r}}
\right)$, where $\hat {\chi} \left( {\vec {r}} \right)$ is
the susceptibility.

For $\hat {\chi}  < < 1$, equation (\ref{eq5}) can be rewritten as

\begin{equation}
\label{eq6}
\Delta \vec {E}\left( {\vec {r},t} \right) -
\frac{{1}}{{c^{2}}}\frac{{\partial ^{2}}}{{\partial t^{2}}}\smallint \hat
{\varepsilon} \left( {\vec {r},t - t^{\prime} } \right)\vec {E}\left( {\vec
{r},t^{\prime} } \right)dt^{\prime}  = 4\pi \left(
{\frac{{1}}{{c^{2}}}\frac{{\partial \vec {j}\left( {\vec {r},t}
\right)}}{{\partial t}} + \vec {\nabla} \rho \left( {\vec {r},t} \right)}
\right).
\end{equation}

In the general case, the susceptibility of the photonic crystal
reads
 $\hat {\chi} \left( {\vec {r}} \right) = \sum\limits_{i} {\hat {\chi
}_{cell} \left( {\vec {r} - \vec {r}_{i}}  \right)} ,$ where $\hat
{\chi }_{cell} \left( {\vec {r} - \vec {r}_{i}}  \right)$ is the
susceptibility of the crystal unit cell. The susceptibility of an
infinite perfect crystal $\hat {\chi} \left( {\vec {r}} \right)$
can be expanded into the Fourier series as follows: $\hat {\chi}
\left( {\vec {r}} \right) = \sum\limits_{\vec {\tau} } {\hat
{\chi} _{\vec {\tau} } e^{i\vec {\tau} \vec {r}}} ,$ where $\vec
{\tau} $ is the reciprocal lattice vector of the crystal.

To be more specific, let us consider in details a practically
important case when a photonic crystal is placed inside a smooth
 waveguide of rectangular cross-section.

The eigenfunctions and the eigenvalues of such a waveguide are
well-studied \cite{15,16}.

Suppose the $z$-axis to be directed along the waveguide axis. Make
the Fourier transform of (\ref{eq5}) over time and longitudinal
coordinate $z$ . Expanding thus obtained equation for the field
$\vec {E}\left( {\vec {r}_{ \bot}  ,k_{z} ,\omega}  \right)$ over
a full set of vector eigenfunctions of a rectangular waveguide
$\vec {Y}^{\lambda }_{mn} \left( {\vec {r}_{ \bot} ,k_{z}}
\right)$ (where $m,n = 1,2,3....$, while $\lambda $ describes the
type of the wave \cite{17},  one can obtain for the field $\vec
{E}$ the equality

\begin{equation}
\label{eq7}
\vec {E}\left( {\vec {r}_{ \bot}  ,k_{z} ,\omega}  \right) =
\sum\limits_{mn\lambda}  {C^{\lambda} _{mn} \left( {k_{z} ,\omega}
\right)\vec {Y}^{\lambda} _{mn}}  \left( {\vec {r}_{ \bot}  ,k_{z}}
\right).
\end{equation}

As a result, the following equation can be written

\begin{equation}
\label{eq8}
\begin{array}{l}
 \left[ {\left( {k_{z} ^{2} + \kappa _{mn\lambda}  ^{2}} \right) -
\frac{{\omega ^{2}}}{{c^{2}}}} \right]C^{\lambda} _{mn} \left( {k_{z}
,\omega}  \right) - \\
 - \frac{{\omega ^{2}}}{{c^{2}}}\,\frac{{1}}{{2\pi
}}\sum\limits_{m'n'\lambda '} {\int {\vec {Y}^{\lambda ^{\ast} }_{mn} \left(
{\vec {r}_{ \bot}  ,k_{z}}  \right)\hat {\chi} \left( {\vec {r}} \right)\vec
{Y}^{\lambda '}_{m'n'} \left( {\vec {r}_{ \bot}  ,k'_{z}}  \right)e^{ -
i\left( {k_{z} - k'_{z}}  \right)z}} \,} d^{2}r_{ \bot}  C_{m'n'}^{\lambda
'} \left( {k'_{z} ,\omega}  \right)dk'_{z} dz = \\
 = \frac{{4\pi i\omega} }{{c^{2}}}\int {\vec {Y}^{\lambda ^{\ast} }_{mn}
\left( {\vec {r}_{ \bot}  ,k_{z}}  \right)\left\{ {\vec {j}\left( {\vec
{r}_{ \bot}  ,z,\omega}  \right) + \frac{{c^{2}}}{{\omega ^{2}}}\vec {\nabla
}\left( {\vec {\nabla} \vec {j}\left( {\vec {r}_{ \bot}  ,z,\omega}
\right)} \right)} \right\}e^{ - ik_{z} z}} d^{2}r_{ \bot}  dz \\
 \end{array}
\end{equation}
\noindent where $\kappa _{mn\lambda}  ^{2} = k_{xm\lambda} ^{2} +
k_{yn\lambda} ^{2} $.

The beam current and density appearing on the right-hand side of
(\ref{eq8}) are complicated functions of the field $\vec {E}$. To
study the problem of the system instability, it is sufficient to
consider the system in the approximation linear  over
perturbation, i.e., one can expand the expressions for $\vec {j}$
and $\rho $ over the field amplitude $\vec {E}$ and abridge
oneself with the linear approximation.

As a result, a closed system of equations comes out. For further
consideration, one should obtain the expressions for the
corrections $\delta \vec {j}$ and $\delta  \rho $ due to beam
perturbation by the field. Considering the Fourier transforms of
the current density and the beam charge $\vec {j}\left( {\vec
{k},\omega} \right)$ and $\rho \left( {\vec {k},\omega} \right)$,
 one can obtain from (\ref{eq2}) that

\begin{equation}
\delta \vec {j}\left( {\vec {k},\omega}  \right) =
e\sum\limits_{\alpha = 1}^{N} {e^{ - i\vec {k}\vec {r}_{\alpha
_{0}} } } \left\{ {\delta \vec {v}_{\alpha}  \left( {\omega - \vec
{k}\vec {u}_{\alpha} }  \right) + \vec {u}_{\alpha}  \frac{{\vec
{k}\delta \vec {v}_{\alpha}  \left( {\omega - \vec {k}\vec
{u}_{\alpha} }  \right)}}{{\omega - \vec {k}\vec {u}_{\alpha} } }}
\right\},
\end{equation}

\noindent
where $\vec {r}_{\alpha _{0}}  $ is the original
coordinate of the electron, $\vec
{u}_{\alpha}  $ is the unperturbed velocity of the electron.

For simplicity, let us consider a cold beam, for which $\vec
{u}_{\alpha}  \approx \vec {u}$, where $\vec {u}$ is the mean
velocity of the beam. The general case of a hot beam is obtained
by averaging $\delta \vec {j}\left({\vec {k},\omega}  \right)$
over the velocity $\vec {u}_{\alpha}$ distribution in the beam.

According to (\ref{eq3}), the velocity $\delta \vec {v}_{\alpha} $
is determined by the field $\vec {E}\left( {\vec {r}_{\alpha}
,\omega}  \right)$ taken at the electron location point  $\vec
{r}_{\alpha}$. The Fourier transform of the field $\vec {E}\left(
{\vec {r}_{\alpha}  ,\omega}  \right)$ has a form

\[
\vec {E}\left( {\vec {r}_{\alpha}  ,\omega}  \right) =
\frac{{1}}{{\left( {2\pi}  \right)^{3}}}\int {\vec {E}\left( {\vec
{k}',\omega} \right)e^{i\vec {k}'\vec {r}_{\alpha} } d^{3}k'}.
\]

As a result, the formula for $\delta \vec {j}\left( {\vec
{k},\omega} \right)$ includes the sum $\sum\limits_{\alpha}  {e^{
- i\left( {\vec {k} - \vec {k}'} \right)\,\vec {r}_{\alpha} } } $
over the particle distribution in the beam. Suppose that the
electrons in an unperturbed beam are uniformly distributed over
the area occupied by the beam. Therefore

\[
\sum\limits_{\alpha}  {e^{ - i\left( {\vec {k} - \vec {k}'} \right)\,\vec
{r}_{\alpha} } } = \left( {2\pi}  \right)^{3}\rho _{0} \,\delta \left( {\vec
{k} - \vec {k}'} \right),
\]

\noindent
where $\rho _{0} $ is the beam
density (the number of electrons per 1 cm$^{3}$).

As a result, the following expression for $\delta \vec {j}\left(
{\vec {k},\omega} \right)$ can be obtained \cite{18,19}:

\begin{equation}
\label{eq9} \delta \vec {j}\left( {\vec {k},\omega}  \right) =
\frac{{i\vec {u}e^{2}\rho \left( {k^{2} - \frac{{\omega
^{2}}}{{c^{2}}}} \right)}}{{\left( {\omega - \vec {k}\vec {u}}
\right)^{2}m\gamma \omega} }\vec {u}\vec {E}\left( {\vec
{k},\omega}  \right).
\end{equation}

Using the continuity equation, one immediately obtains the
expression for $\rho \left( {\vec {k},\omega}  \right)$.
Expression (\ref{eq9}), the inverse Fourier transform of $\vec
{E}\left( {\vec {k},\omega}  \right)$, and the expansion
(\ref{eq7}) enable writing the system of equations (\ref{eq8}) as
follows:
\begin{equation}
\label{eq10}
\begin{array}{l}
 \left[ {\left( {k_{z} ^{2} + \kappa _{mn\lambda}  ^{2}} \right) -
\frac{{\omega ^{2}}}{{c^{2}}}} \right]C^{\lambda} _{mn} \left( {k_{z}
,\omega}  \right) - \\
 - \frac{{\omega ^{2}}}{{c^{2}}}\,\frac{{1}}{{2\pi
}}\sum\limits_{m'n'\lambda '} {\int {\vec {Y}^{\lambda ^{\ast} }_{mn} \left(
{\vec {r}_{ \bot}  ,k_{z}}  \right)\hat {\chi} \left( {\vec {r}} \right)\vec
{Y}^{\lambda '}_{m'n'} \left( {\vec {r}_{ \bot}  ,k'_{z}}  \right)e^{ -
i\left( {k_{z} - k'_{z}}  \right)z}} \,} d^{2}r_{ \bot}  C_{m'n'}^{\lambda
'} \left( {k'_{z} ,\omega}  \right)dk'_{z} dz = \\
 = - \frac{{\omega _{L}^{2} \left( {k_{mn}^{2} c^{2} - \omega ^{2}}
\right)}}{{\gamma c^{4}\left( {\omega - \vec {k}_{mn} \vec {u}}
\right)^{2}}}\left\{ {\frac{{1}}{{2\pi} }\left| {\int {\vec {u}\,\vec
{Y}^{\lambda} _{mn} \left( {\vec {k}_{ \bot}  ,k_{z}}  \right)d^{2}k_{ \bot
}} }  \right|^{2}} \right\}C^{\lambda} _{mn} \left( {k_{z} ,\omega}
\right), \\
 \end{array}
\end{equation}

\noindent where $\vec {Y}^{\lambda} _{mn} \left( {\vec {k}_{ \bot}
,k_{z}}  \right) = \int {e^{ - i\vec {k}_{ \bot}  \vec {r}_{ \bot}
} \vec {Y}_{mn}^{\lambda} } \left( {\vec {r}_{ \bot}  ,k_{z}}
\right)d^{2}r_{ \bot}  $.

Note that within the limit where the transverse dimensions of a
photonic crystal tend to infinity, the expression between the
braces takes the form $\left( {\vec {e}\vec {u}} \right)^{2}$,
where $\vec{e}$  is the unit polarization vector of the wave
emitted by the beam \cite{18,19}.

Now let us consider the integrals on the left-hand side of
equation (\ref{eq10}). Note that according to \cite{15,16,17}, the
eigenfunctions $\vec {Y}_{mn}^{\lambda} \left({\vec {r}_{ \bot}
,k_{z}}  \right)$ of a rectangular waveguide  include the
combinations of sines and cosines of the form $sin\frac{{\pi
m}}{{a}}x,cos\frac{{\pi m}}{{a}}x$ ($sin\frac{{\pi
n}}{{b}}y,cos\frac{{\pi n}}{{b}}y$), i.e., in fact, the
combinations $e^{i\frac{{\pi m}}{{a}}x},e^{i\frac{{\pi
n}}{{b}}y}$.
Hence, the left-hand side of the equation includes
the integrals of the type

\[
I = \int {e^{ - i\frac{{\pi m}}{{a}}x}} \sum\limits_{i} {\hat {\chi} _{cell}
} \left( {x - x_{i} ,y - y_{i} ,z - z_{i}}  \right)e^{i\frac{{\pi
m'}}{{a}}x}dx.
\]

The substitution of variables $x - x_{i} = \eta $ gives the sums
of the form

\[
S_{x} = \sum\limits_{i} {e^{ - i\frac{{\pi} }{{a}}\left( {m - m'}
\right)x_{i}} }
\]

\noindent where $x_{i} = d_{x} f_{1} $, where $d_{x} $ is the
period of the photonic crystal along the $x$-axis, $f_{1} =
1,2,...N_{x} $, where $N_{x} $ is the number of cells along the
$x$-axis.

The above-mentioned sum

\begin{equation}
\label{eq11} S_{x} = \sum\limits_{i} {e^{ - i\frac{{\pi}
}{{a}}\left( {m - m'} \right)x_{i}} } = e^{i\frac{{\pi}
}{{2a}}\left( {m - m'} \right)\left( {N_{x} - 1} \right)d_{x}}
\frac{{sin  \frac{{\pi \left( {m - m'} \right)d_{x} N_{x} }}{{2a}}
} }{{sin\frac{{\pi \left( {m - m'} \right)d_{x}} }{{2a}}}} .
\end{equation}

If $m - m' = 0$, then $S_{x} = N_{x} $.

 Let us now discuss what this sum is equal to when $m - m' = 1$. In the
 numerator $d_{x} N_{x} = a$,
 hence, the nominator {is equal to} 1 ($sin\frac{{\pi} }{{2}} = 1$), while in the denominator $sin\frac{{\pi
d_{x}} }{{2a}} \approx \frac{{\pi} }{{2N_{x}} }$.
As a result, the relation
$\frac{{S_{x} \left( {m - m' = 1} \right)}}{{S_{x}
\left( {m - m' = 0} \right)}} = \frac{{2}}{{\pi} } \approx 0.6$.

With growing difference $m - m'$, the contribution to the sum of
the next terms diminishes until the
following equality is fulfilled
\begin{equation}
\label{eq12}
\frac{{\pi \left( {m - m'} \right)d_{x}} }{{2a}} = \pi {\rm P},
\end{equation}

\noindent
where ${\rm P} = \pm 1, \pm 2...$ In this case the sum $S_{x} = N_{x} $.

The similar reasoning is valid for summation along the $y$-axis.

It follows from the aforesaid that if the equalities like
(\ref{eq11}), (\ref{eq12}) are fulfilled, that is, the equalities
$k_{xm} - k'_{xm'} = \tau _{x} $ are fulfilled (i.e., $k'_{xm'} =
k_{xm} - \tau _{x} $), where $\tau _{x} = \frac{{2\pi} }{{d_{x}}
}F$ is the $x$-component of the reciprocal lattice vector of the
photonic crystal,  $F = 0, \pm 1, \pm 2...$ and $k_{yn} - k'_{yn'}
= \tau _{y} $ (i.å., $k'_{yn'} = k_{yn} - \tau _{y} $), where
$\tau _{y} = \frac{{2\pi} }{{d_{y}} }F'$ is the $y$-component of
the reciprocal lattice vector of the photonic crystal,  $F' = 0,
\pm 1, \pm2...$), then the major contribution to the sums comes
from the amplitudes $C_{m'n'}^{\lambda '} \left( {k'_{z} ,\omega}
\right) \equiv C^{\lambda '}\left( {\vec {k}_{ \bot mn} - \vec
{\tau} _{ \bot}  ,k_{z} - \tau _{z} ,\omega}  \right) = C^{\lambda
'}\left( {\vec {k}_{mn} - \vec {\tau} ,\omega}  \right)$.

In describing the system we shall further consider only those
modes that satisfy the equalities of the type (\ref{eq11}),
(\ref{eq12}). As stated above, the contribution of other modes is
suppressed.

As a result, one can rewrite the system of equations (\ref{eq10})
as

\begin{equation}
\label{eq13}
\begin{array}{l}
 \left( {\vec {k}_{mn}^{2} - \frac{{\omega ^{2}}}{{c^{2}}}}
\right)C^{\lambda} \left( {\vec {k}_{mn} ,\omega}  \right) - \frac{{\omega
^{2}}}{{c^{2}}}\sum\limits_{\lambda '\tau}  {\chi _{mn}^{\lambda \lambda '}
\left( {\vec {\tau} } \right)} C^{\lambda '}\left( {\vec {k}_{mn} - \vec
{\tau} ,\omega}  \right) = \\
 - \frac{{\omega _{L}^{2} \left( {k_{mn}^{2} c^{2} - \omega ^{2}}
\right)}}{{\gamma c^{4}\left( {\omega - \vec {k}_{mn} \vec {u}}
\right)^{2}}}\left\{ {\frac{{1}}{{2\pi} }\left| {\int {\vec {u}\,\vec
{Y}^{\lambda} _{mn} \left( {\vec {k}_{ \bot}  ,k_{z}}  \right)d^{2}k_{ \bot
}} }  \right|^{2}} \right\}C^{\lambda} \left( {\vec {k}_{mn} ,\omega}
\right), \\
 \end{array}
\end{equation}

\noindent
i.e.,

\begin{equation}
\label{eq14}
\begin{array}{l}
 \left( {\vec {k}_{mn}^{2} - \frac{{\omega ^{2}}}{{c^{2}}}\left( {1 + \chi
_{mn}^{\lambda \lambda}  \left( {0} \right) - \frac{{\omega _{L}^{2} \left(
{k_{mn}^{2} c^{2} - \omega ^{2}} \right)}}{{\omega ^{2}\gamma c^{2}\left(
{\omega - \vec {k}_{mn} \vec {u}} \right)^{2}}}\left\{ {\frac{{1}}{{2\pi
}}\left| {\int {\vec {u}\,\vec {Y}^{\lambda} _{mn} \left( {\vec {k}_{ \bot}
,k_{z}}  \right)d^{2}k_{ \bot} } }  \right|^{2}} \right\}} \right)}
\right)C^{\lambda} \left( {\vec {k}_{mn} ,\omega}  \right) \\
 - \frac{{\omega ^{2}}}{{c^{2}}}\sum\limits_{\lambda '\tau}  {\chi
_{mn}^{\lambda \lambda '} \left( {\vec {\tau} } \right)} C^{\lambda '}\left(
{\vec {k} - \vec {\tau} ,\omega}  \right) = 0 \\
 \end{array}
\end{equation}

\noindent where $\chi _{mn}^{\lambda \lambda '} \left( {\tau}
\right) = \frac{{1}}{{d_{z}} }\int {\vec {Y}^{\lambda \ast} _{mn}
\left( {\vec {r}_{ \bot}  ,k_{z}}  \right)\hat {\chi} \left( {\vec
{r}_{ \bot}  ,\tau _{z}} \right)} \vec {Y}_{m'n'}^{\lambda '}
\left( {\vec {r}_{ \bot}  ,k_{z} - \tau _{z}}  \right)d^{2}r_{
\bot}  $, $\hat {\chi} \left( {\vec {r}_{ \bot} ,\tau _{z}}
\right) = \sum\limits_{x_{i} ,y_{i}}  {\int {\hat {\chi }_{cell}
\left( {x - x_{i} ,y - y_{i} ,\zeta}  \right)}}  \,e^{ - i\tau
_{z} \zeta} d\zeta $, $m'$ and  $n'$ are found by the conditions
like (\ref{eq12}), $\omega _{L} $ is the Langmuir frequency,
$\omega _{_{L}} ^{2} = \frac{{4\pi e^{2}\rho _{0} }}{{m}}$.

This system of equations coincides in form with that describing
the instability of a beam passing through an infinite crystal
\cite{18,19}. The difference between them is that the coefficients
appearing in these equations are defined differently and that in the
case of an infinite crystal, the wave vectors $\vec {k}_{mn} $ have a
continuous spectrum of eigenvalues rather than a discrete one.

These equations enable one to define the dependence $k\left(
{\omega} \right)$, thus defining the expressions for the waves
propagating in the crystal. By matching the incident wave packet
and the set of waves propagating inside the photonic crystal using
the boundary conditions, one can obtain the explicit expression
describing the solution of the considered equations.

The result obtained is formally analogous to that given in \cite{20}.

According to (\ref{eq14}), the expression between
the square brackets acts as the dielectric permittivity $\varepsilon $ of
the crystal under the conditions when diffraction can be neglected:
\[\varepsilon _{0} = n^{2} = 1 + \chi _{mn}^{\lambda \lambda}
\left( {0} \right) - \frac{{\omega _{L}^{2} \left( {k_{mn}^{2} c^{2} -
\omega ^{2}} \right)}}{{\omega ^{2}\gamma c^{2}\left( {\omega - \vec
{k}_{mn} \vec {u}} \right)^{2}}}\left\{ {\frac{{1}}{{2\pi} }\left| {\int
{\vec {u}\,\vec {Y}^{\lambda} _{mn} \left( {\vec {k}_{ \bot}  ,k_{z}}
\right)d^{2}k_{ \bot} } }  \right|^{2}} \right\},
\]
 $n $  is the refractive index.

As is seen, in this case the contribution to the refractive index comes
 not only from the scattering of waves by the unit cell of the crystal
 lattice, but also from the scattering of waves by the beam electrons
 (the term proportional to $\omega _{L}^{2} $): the photonic crystal
 penetrated by a beam of electrons is a medium that can be described
  by a ceratin refractive index $n$ (or the dielectric permittivity
  $\varepsilon _{0} $).

According to (\ref{eq14}), the beam contribution increases when
$\omega \to \vec {k}\vec {u}$.

Since this system of equations is homogeneous, its solvability
condition is the vanishing of the system determinant.

In the beginning, let us assume that the diffraction conditions
are not fulfilled. Then the amplitudes of diffracted waves are
small. In this case the sum over $\tau $ can be dropped, and the
conditions for the occurrence of the wave in the system is
obtained by the requirement that the expression between the square
brackets equal zero.

This expression can be written in the form (the velocity $\vec {u}
|| oz$)

\[
\left( {\omega - k_{z} u} \right)^{2}\left( {k_{mn}^{2} - \frac{{\omega
^{2}}}{{c^{2}}}n_{0}^{2}}  \right) = - \frac{{\omega _{L}^{2} \left(
{k_{mn}^{2} c^{2} - \omega ^{2}} \right)}}{{\gamma c^{4}}}\left\{
{\frac{{1}}{{2\pi} }\left| {\int {\vec {u}\,\vec {Y}^{\lambda} _{mn} \left(
{\vec {k}_{ \bot}  ,k_{z}}  \right)d^{2}k_{ \bot} } }  \right|^{2}}
\right\},
\]

\noindent
where  $n_{0} $ is the refractive index of the
photonic crystal in the absence of the beam
$\varepsilon _{0} = n_{0}^{2} = 1 + \chi _{mn}^{\lambda \lambda}  \left( {0}
\right)$,

\noindent
i.e.,

\begin{equation}
\label{eq15} \left( {k_{z}^{2} - \left( {\frac{{\omega
^{2}}}{{c^{2}}}n_{0}^{2} - \kappa_{mn}^{2}}  \right)}
\right)\left( {\omega - k_{z} u} \right)^{2} = - \frac{{\omega
_{L}^{2} \left( {k_{mn}^{2} c^{2} - \omega ^{2}} \right)}}{{\gamma
c^{4}}}\left\{ {\frac{{1}}{{2\pi} }\left| {\int {\vec {u}\,\vec
{Y}^{\lambda} _{mn} \left( {\vec {k}_{ \bot}  ,k_{z}}
\right)d^{2}k_{ \bot} } }  \right|^{2}} \right\}
\end{equation}

Since the nonlinearity is insignificant, let us consider as the zero
 approximation the spectrum of the waves of equation (\ref{eq15}) with
  zero right-hand side.

Let us concern with the case when $\omega - k_{z} u \to 0$ (i.e.,
the Cherenkov radiation condition can be fulfilled) and $\left(
{k_{z}^{2} - \left( {\frac{{\omega ^{2}}}{{c^{2}}}n_{0}^{2} -
\kappa _{mn}^{2}}  \right)} \right) \to 0$, i.e, the
electromagnetic wave can propagate in a photonic crystal without
the beam. With zero right-hand side the equation reads

\begin{equation}
\label{eq16}
\left( {k_{z}^{2} - \left( {\frac{{\omega ^{2}}}{{c^{2}}}n_{0}^{2} - \kappa
_{mn}^{2}}  \right)} \right) = 0,
\quad
\left( {\omega - k_{z} u} \right) = 0
\end{equation}

As a consequence, in this case the roots of the equation are
\begin{equation}
\label{eq17} k_{1z} = \frac{{\omega} }{{c}}\sqrt {n_{0}^{2} -
\frac{{\kappa _{mn}^{2} c^{2}}}{{\omega ^{2}}}} , \quad k'_{1z} =
- k_{1z} , \quad k_{2z} = \frac{{\omega} }{{u}}.
\end{equation}

Since  $k_{2z} = \frac{{\omega} }{{u}}
> 0$ in view of the Cherenkov condition, we are concerned with the propagation of the
wave with $k_{1z} > 0$ in the photonic crystal. In this case in
the equation for $k_{z} $, one can take $\left( {k_{z} - k_{1z}}
\right) \left( {k_{z} + k_{1z}}  \right) \approx 2k_{1z} \left(
{k_{z} - k_{1z}} \right)$ and rewrite equation (\ref{eq15}) as
follows:

\begin{equation}
\label{eq18}
\left( {k_{z} - k_{1z}}  \right)\left( {k_{z} - k_{2z}}  \right)^{2} = -
\frac{{\omega _{L}^{2} \omega ^{2}\left( {n_{0}^{2} - 1} \right)}}{{2k_{1z}
u^{2}\gamma c^{4}}}\left\{ {\frac{{1}}{{2\pi} }\left| {\int {\vec {u}\,\vec
{Y}^{\lambda} _{mn} \left( {\vec {k}_{ \bot}  ,k_{z}}  \right)d^{2}k_{ \bot
}} }  \right|^{2}} \right\}
\end{equation}

\noindent
i.e.,

\begin{equation}
\label{eq19}
\left( {k_{z} - k_{1z}}  \right)\left( {k_{z} - k_{2z}}  \right)^{2} = - A
\end{equation}

\noindent

where $A$ is real and $A > 0$ (as for the occurrence of the
Cherenkov effect, it is necessary that $n_{0}^{2} > 1$).
We have obtained the  cubic equation for  $k_{z} $.
Let us consider the case when the roots $k_{1z} $
and  $k_{2z} $ coincide $k_{1z} = k_{2z} $. It is possible
when the particle velocity satisfies the condition

\begin{equation}
\label{eq20} u = \frac{{c}}{{\sqrt {n_{0}^{2} - \frac{{\kappa
_{mn}^{2} c^{2}}}{{\omega ^{2}}}}} }.
\end{equation}

\noindent Introduction of  $\xi = k - k_{1z} $ gives for $k_{1z} =
k_{2z} $

\begin{equation}
\label{eq21} \xi ^{3} = - A.
\end{equation}

\noindent The solution of equation (\ref{eq21}) gives three roots
$\xi _{1} = - \sqrt[{3}]{{A}}$,
 $\xi
_{2,3} = \frac{{1}}{{2}}\left( {1 \pm i\sqrt {3}}  \right)
\sqrt[{3}]{{A}}$.

As a consequence, the state corresponding to the root $\xi _{2} =
\frac{{1}}{{2}}\left( {1 + i\sqrt {3}}  \right)\sqrt[{3}]{{A}}$
grows with growing $z$, which indicates the presence of
instability in a beam \cite{21}. In this case $Im\,k_{z} = Im\,\xi
_{2}\sim\sqrt[{3}]{{\rho} }$.

Note here that the photonic crystal built from metallic threads
has the refractive index $n_{0} < 1$ for a wave with the electric
polarizability parallel to the threads, i.e., in this case the
Cherenkov instability of the beam does not exist \cite{9} (but if
the electric vector of the wave is orthogonal to the metallic
threads, the refractive index is  $n_{0} > 1$ , so for such a wave
the Cherenkov instability exists \cite{22}).

It should be pointed out, however, that, unlike an infinite
photonic crystal, the field in the crystal placed into the
waveguide has a mode character, and so the presence of $\kappa
_{mn}^{2} $ in the denominator of equation (\ref{eq20}) results in
reduction  of the radicand in (\ref{eq20}) to the magnitude
smaller than unity even when $n_{0}^{2}
> 1$. Hence, $u > c$, which is impossible.
Consequently, the radiative instability of the above type in the
waveguide can arise under the condition $n_{0}^{2} - \frac{{\kappa
_{mn}^{2} c^{2}}}{{\omega ^{2}}} >1$ rather than $n_{0}^{2}>1$.

Suppose now that in the photonic crystal the conditions can be
realized under which the wave amplitude $C_{mn} \left(
{\vec{k}_{mn} + \vec {\tau} } \right)$ is comparable with the
amplitude $C_{mn} \left( {\vec {k}_{mn}} \right)$. By analogy with
the standard diffraction theory for an infinite crystal \cite{23,
24}, in the case under consideration, when $\chi < < 1$, it is
sufficient that only the equations for these amplitudes remain in
(\ref{eq14}).

To be specific, let us further consider a photonic crystal formed
by parallel threads. Also assume that they are parallel to the
waveguide boundary $\left( {y,z} \right)$.

Analysis of diffraction of  a $\lambda $-type wave with the
electric vector in the plane $\left( {y,z} \right)$ (a TM-wave)
gives

\begin{equation}
\label{eq22}
\left[ {k_{mn}^{2} - \frac{{\omega ^{2}}}{{c^{2}}}\varepsilon}
\right]C^{\lambda} \left( {\vec {k}_{mn} ,\omega}  \right) - \frac{{\omega
^{2}}}{{c^{2}}}\chi _{mn}^{\lambda \lambda}  \left( { - \vec {\tau} }
\right)C^{\lambda} \left( {\vec {k}_{mn} + \vec {\tau} ,\omega}  \right) =
0
\end{equation}

\[
\left[ {\left( {\vec {k}_{mn} + \vec {\tau} } \right) - \frac{{\omega
^{2}}}{{c^{2}}}\varepsilon _{0}}  \right]C^{\lambda} \left( {\vec {k}_{mn} +
\vec {\tau} ,\omega}  \right) - \frac{{\omega ^{2}}}{{c^{2}}}\chi
_{mn}^{\lambda \lambda}  \left( {\vec {\tau} } \right)C^{\lambda} \left(
{\vec {k}_{mn} ,\omega}  \right) = 0.
\]

Since the term containing $\left( {\omega - \left( {\vec {k} +
\vec {\tau} } \right)\vec {u}} \right)^{ - 1}$ is small when
$\left( {\omega - \vec {k}\vec {u}} \right)$ vanishes, in the
second equation it is dropped.

The dispersion equation defining the relation between $k_{z} $ and
$\omega $ is obtained by equating to zero the determinant of the
system (\ref{eq22}) and has a form:

\begin{equation}
\label{eq23}
\begin{array}{l}
 \left[ {\left( {k_{mn}^{2} - \frac{{\omega ^{2}}}{{c^{2}}}\varepsilon _{0}
} \right)\left( {\left( {\vec {k}_{mn} + \vec {\tau} } \right)^{2} -
\frac{{\omega ^{2}}}{{c^{2}}}\varepsilon _{0}}  \right) - \frac{{\omega
^{4}}}{{c^{4}}}\chi _{\tau}  \chi _{ - \tau} }  \right]\left( {\omega -
k_{z} u} \right)^{2} = \\
 - \frac{{\omega _{L}^{2}} }{{\gamma c^{4}}}\left\{ {\frac{{1}}{{2\pi
}}\left| {\int {\vec {u}\,\vec {Y}^{\lambda} _{mn} \left( {\vec
{k}_{ \bot} ,k_{z}}  \right)d^{2}k_{ \bot} } }  \right|^{2}}
\right\}\left( {k_{mn}^{2} c^{2} - \omega ^{2}} \right)\left(
{\left( {\vec {k}_{mn} + \vec {\tau} } \right)^{2} - \frac{{\omega
^{2}}}{{c^{2}}}\varepsilon _{0}}  \right).
 \end{array}
\end{equation}

Because the right-hand side of the equation is small, one can
again seek the solution near the points where the right-hand side
is zero that corresponds the condition of occurrence of the
Cherenkov radiation and excitation of the wave which can propagate
in the waveguide:

\begin{equation}
\label{eq24}
\begin{array}{l}
 \left( {k_{z}^{2} - \left( {\frac{{\omega ^{2}}}{{c^{2}}}\varepsilon _{0} -
\kappa _{mn}^{2}}  \right)} \right)\left( {\left( {k_{z} + \tau}
\right)^{2} - \left( {\frac{{\omega ^{2}}}{{c^{2}}}\varepsilon _{0} - \left(
{\vec {\kappa} _{mn} + \vec {\tau} _{ \bot} }  \right)^{2}} \right)} \right)
- \frac{{\omega ^{4}}}{{c^{4}}}\chi _{\tau}  \chi _{ - \tau}  = 0 \\
 \left( {k_{z} - \frac{{\omega} }{{u}}} \right)^{2} = 0 \\
 \end{array}
\end{equation}

The roots of equations are sought near the conditions
$k_{mn}^{2} \approx \left( {\vec
{k}_{mn} + \vec {\tau} } \right)$,

\begin{equation}
\label{eq25}
k_{z} = k_{z0} + \xi ,
\quad
k_{z}^{2} = k_{z0}^{2} + 2k_{z0} \xi + \xi ^{2},
\quad
k_{z0}^{2} = \frac{{\omega ^{2}}}{{c^{2}}}\varepsilon _{0} - \kappa
_{mn}^{2} ;
\quad
k_{z0} = \frac{{\omega} }{{c}}\sqrt {\varepsilon _{0} - \frac{{\kappa
_{mn}^{2} c^{2}}}{{\omega ^{2}}}}
\end{equation}

\[
\left( {k_{z} + \tau _{z}}  \right)^{2} = \left[ {\left( {k_{0z} + \tau _{z}
} \right) + \xi}  \right]^{2} = \left( {k_{0z} + \tau _{z}}  \right)^{2} +
2\left( {k_{0z} + \tau _{z}}  \right)\xi + \xi ^{2}
\]

Hence,

\begin{equation}
\label{eq26}
\begin{array}{l}
 \left( {k_{0z} + \tau _{z}}  \right)^{2} + \left( {\vec {\kappa} _{mn} +
\vec {\tau} _{ \bot} }  \right)^{2} + 2\left( {k_{0z} + \tau _{z}}
\right)
+ 2\left( {k_{0z} + \tau _{z}}  \right)\xi + \xi ^{2} = \\
 \left( {\vec {k}_{mn} + \vec {\tau} } \right)^{2} + 2\left( {k_{0z} + \tau
_{z}}  \right)\xi + \xi ^{2} = k_{0mn}^{2} + 2\vec {k}_{0mn} \vec
{\tau} + \tau ^{2} + 2\left( {k_{0z} + \tau _{z}}  \right)\xi +
\xi ^{2}.
 \end{array}
\end{equation}

And one can get

\[
2k_{0z} \xi \left( {2\left( {k_{0z} + \tau _{z}}  \right)\xi + \left( {2\vec
{k}_{0mn} \vec {\tau}  + \tau ^{2}} \right)} \right) - \frac{{\omega
^{4}}}{{c^{4}}}\chi _{\tau}  \chi _{ - \tau}  = 0
\]

\begin{equation}
\label{eq27}
4k_{0z} \left( {k_{0z} + \tau _{z}}  \right)\xi ^{2} + 2k_{0z} \left( {2\vec
{k}_{0mn} \vec {\tau}  + \tau ^{2}} \right)\xi - \frac{{\omega
^{4}}}{{c^{4}}}\chi _{\tau}  \chi _{ - \tau}  = 0
\end{equation}

\[
\xi ^{2} + \frac{{\left( {2\vec {k}_{0mn} \vec {\tau}  + \tau ^{2}}
\right)}}{{\left( {k_{0z} + \tau _{z}}  \right)}}\xi - \frac{{\omega
^{4}}}{{c^{4}}}\frac{{\chi _{\tau}  \chi _{ - \tau} } }{{4k_{0z} \left(
{k_{0z} + \tau _{z}}  \right)}} = 0
\]

\[
\xi _{1,2} = - \frac{{\left( {2\vec {k}_{0} \vec {\tau}  + \tau ^{2}}
\right)}}{{4\left( {k_{0z} + \tau _{z}}  \right)}} \pm \sqrt {\left(
{\frac{{2\vec {k}_{0} \vec {\tau}  + \tau ^{2}}}{{4\left( {k_{0z} + \tau
_{z}}  \right)}}} \right)^{2} + \frac{{\omega ^{4}}}{{c^{4}}}\frac{{\chi
_{\tau}  \chi _{ - \tau} } }{{4k_{0z} \left( {k_{0z} + \tau _{z}}
\right)}}}
\]

If $\left( {k_{0z} + \tau _{z}}  \right) = - \left| {k_{0z} + \tau _{z}}
\right|$, the root can cross the zero point. At the same time,
the second equation should hold.

\[
\omega - k_{z} u = \omega - k_{0z} u - \xi u = 0.
\]

Consequently,

\[
\xi = \frac{{\omega - k_{0z} u}}{{u}} = \frac{{\omega} }{{u}} -
k_{0z} = \frac{{\omega} }{{u}} - \frac{{\omega} }{{c}}\sqrt
{\varepsilon _{0} - \frac{{\kappa _{mn}^{2} c^{2}}}{{\omega
^{2}}}}.
\]

If $\varepsilon _{0} < 1$, then $\xi = \frac{{\omega} }{{u}}\left( {1 -
\beta \sqrt {\varepsilon _{0} - \frac{{\kappa _{mn}^{2} c^{2}}}{{\omega
^{2}}}}}  \right) > 0$, $\xi = \frac{{\omega} }{{u}} - k_{0z} $

Let the roots  $\xi _{1} $ and $\xi _{2} $ coincide ($\xi _{1} = \xi _{2} $).
This is possible at point

\[
\frac{{2\vec {k}_{0} \vec {\tau}  + \tau ^{2}}}{{4\left( {k_{0z} + \tau _{z}
} \right)}} = \pm \frac{{\omega ^{2}}}{{c^{2}}}\frac{{\sqrt {\chi _{\tau}
\chi _{ - \tau} } } }{{\sqrt {4k_{0z} \left| {k_{0z} + \tau _{z}}  \right|}
}},
\]

\noindent
here $k_{0z} + \tau _{z} < 0$.

The roots coincide when the following equality is fulfilled

 \[\frac{{\omega} }{{u}} - k_{0z} = \mp \frac{{\omega
^{2}}}{{c^{2}}}\frac{{\sqrt {\chi _{\tau}  \chi _{ - \tau} } } }{{\sqrt
{4k_{0z} \left| {k_{0z} + \tau _{z}}  \right|}} },
\]
 i.e.,
 \[\frac{{\omega} }{{u}}
= k_{0z} \mp \frac{{\omega ^{2}}}{{c^{2}}}\frac{{\sqrt {\chi _{\tau}  \chi
_{ - \tau} } } }{{\sqrt {4k_{0z} \left| {k_{0z} + \tau _{z}}  \right|}} } \quad\mbox{and}\quad
k_{0z} = \frac{{\omega} }{{c}}\sqrt {\varepsilon _{0} - \frac{{\kappa
_{mn}^{2} c^{2}}}{{\omega ^{2}}}}
\]

Let $\varepsilon _{0} < 1$, then $\frac{{\omega} }{{u}} > k_{0z} $
(since $u < c$),  the situation for the solution $\frac{{\omega}
}{{u}} = k_{0z} - \frac{{\omega ^{2}}}{{c^{2}}}\frac{{\sqrt {\chi
_{\tau}  \chi _{ - \tau} } } }{{\sqrt {4k_{0z} \left| {k_{0z} +
\tau _{z}}  \right|}} }$ gets complicated and the
Vavilov-Cherenkov condition is not fulfilled.

Now let us consider the solution $\frac{{\omega} }{{u}} = k_{0z} +
\frac{{\omega ^{2}}}{{c^{2}}}\frac{{\sqrt {\chi _{\tau}  \chi _{ -
\tau} } } }{{\sqrt {4k_{0z} \left| {k_{0z} + \tau _{z}}  \right|}}
}$. At $\tau _{z} < 0$ the difference $k_{0z} + \tau _{z} $ can be
reduced so that the sum on the right would appear to become equal
to  $\frac{{\omega} }{{u}}$, and so  one could
 obtain 4 coinciding roots.

Interestingly enough, for backward diffraction, which is a typical
case of frequently used one-dimensional generators with a
corrugated metal waveguide (the traveling-wave tube, the
backward-wave tube), such a coincidence of roots is impossible.

Indeed, let the roots $\xi _{1} $ and $\xi _{2} $ coincide. In
this case for the backward Bragg diffraction
 $\left| {\tau _{z}}  \right| \approx
2k_{0z} $, $\tau _{z} < 0$. Then by substituting the expressions
for  $k_{0z} = \frac{{\omega} }{{c}}\sqrt {\varepsilon _{0} -
\frac{{\kappa _{mn}^{2} c^{2}}}{{\omega ^{2}}}} $ and $\varepsilon
_{0} = n_{0}^{2} = 1 + \chi _{mn}^{\lambda \lambda}  \left( {0}
\right)$ and retaining the first-order infinitesimal terms, the
relation $\frac{{\omega} }{{u}} \approx k_{0z} + \frac{{\omega
^{2}}}{{c^{2}}}\frac{{\left| {\chi _{\tau} } \right|}}{{2k_{0z}}
}$ can be reduced to the form $\frac{{\omega} }{{u}} \approx
\frac{{\omega }}{{c}}\left( {1 - \frac{{\left| {\chi
_{mn}^{\lambda \lambda}  \left( {0} \right)} \right|}}{{2}} -
\frac{{\kappa _{mn}^{2} c^{2}}}{{2\omega ^{2}}} + \frac{{\omega}
}{{c}}\frac{{\left| {\chi _{\tau} }  \right|}}{{2}}} \right) <
\frac{{\omega} }{{u}}$, i.e., the equality does not hold and the
four-fold degeneracy is impossible. Only ordinary three-fold
degeneration is possible.

However, if $\varepsilon_{0}>1$ and is appreciably large,  then in
a one-dimensional case, the four-fold degeneracy of roots is also
possible in a finite photonic crystal\footnote{The authors are
grateful to K. Batrakov, who drew our attention to the fact that
for an infinite crystal with $\varepsilon_{0}>1$, the intersection
of roots is possible in a one-dimensional case.}.

Thus, the left-hand side of equation (\ref{eq23}) has  four roots
$\xi_{1}$, $\xi_{2}$, and a double degenerated root $\xi_{3}$.
Hence, equation (\ref{eq23}) can be written as follows:

\[
\left( {\xi - \xi _{1}}  \right)\left( {\xi - \xi _{2}}  \right)
\left( {\xi - \xi _{3}}  \right)^{2} = B.
\]

If the roots coincide ($\xi _{1} = \xi _{2} = \xi _{3} $), one
obtains
 $\left( {\xi - \xi _{1}}  \right)^{4} = B,$, i.e., $\xi - \xi _{1} =
\sqrt[{4}]{{B}}$.

The fourth root of  $B$ has imaginary solutions depending on the
beam density as
 $Imk_{z}\sim\rho _{0} ^{1/4}$ (the parameter
$B \sim \omega _{L}^{2} $, i.e., $B \sim \rho _{0} $,  see the
right-hand side of (\ref{eq23})). This increment is larger than
the one we obtained for the case of the three-fold degeneracy.

The analysis shows that with increasing number of diffracted
waves, the law established in \cite{6,7,8} is valid: the
instability increment appears to be proportional to $\rho
^{\frac{{1}}{{s + 3}}}$, where $s$ is the number of waves emerging
through diffraction. As a result,  the abrupt decrease in the
threshold generation current also remains in this case (the
threshold generation current $j_{th}\sim\frac{{1}}{{\left( {kL}
\right)^{3}\left( {k\chi _{\tau}  L} \right)^{2s}}}$, where $L$ is
the length of the interaction area).

It is interesting that according to \cite{12}, for a photonic
crystal made from metallic threads, the coefficients $\chi \left(
{\tau}  \right)$, defining the threshold current and the growth of
the beam instability, are practically independent on $\tau $ up to
the terahertz range of frequencies because the diameter of the
thread can easily be made much smaller than the wavelength.
That is why photonic crystals with the period of about 1 mm can
 be used for lasing in terahertz range at high harmonics (for
example,  photonic crystal with 3 mm period provides the frequency
of the tenth harmonic of about 1 terahertz ($\lambda $=300
micron).

The analysis of laser generation in VFEL with a photonic crystal
 when the beam moves in an undulator (electromagnetic wave)
located in a finite crystal, made similarly to the above analysis,
shows that in this case the dispersion equation and the law of
instability also have the same form as in the case of an infinite
crystal. The procedure for going from  the dispersion equations
describing instability in the infinite case (\ref{eq14}) is
similar to that discussed earlier in this paper. It consists in
replacing the continuous $\vec {k}$ by the quantified values of
$\vec {k}_{mn} $  and redefining the coefficients appearing in
equations like (\ref{eq14}).

It is important to emphasize the general character of the rules
found in this paper for obtaining dispersion equations that
describe the radiative instability of the electron beam in a
finite photonic crystal. In particular, they are valid for
describing the processes of instability of an electromagnetic wave
in finite nonlinear photonic crystals.

\end{document}